\documentclass[preprint,authoryear,12pt]{elsarticle}

\usepackage{graphicx}
\usepackage{amssymb}

\journal{Planetary and Space Science (Short communication) \, }

\begin{document}

\begin{frontmatter}

\title{Modulation of the Solar Wind Velocity by Mercury}

\author{I. F. Nikulin\corref{lbl:corauth}}

\cortext[lbl:corauth]{Corresponding author.
Tel:~+7 495 939 19 73
\ead{ifn@sai.msu.ru}}

\address{M.V. Lomonosov Moscow State University,
         Sternberg Astronomical Institute
         Universitetski pr. 13,
         119991 Moscow, Russia}

\begin{abstract}
To study the variations in the solar wind velocity during inferior
conjunctions of Mercury and Earth, we analyzed 54 events in the period
1995 to 2012 by the superimposed epoch method. We have found
a noticeable increase in the velocity both before and after
the conjunctions as well as decrease in the velocity within 3--4 days
after them, which seems to be associated with Mercury's ``shadow''.
The results obtained might be used to improve a forecast of
the solar wind velocity.
\end{abstract}

\begin{keyword}
Solar wind \sep Mercury \sep magnetospheres
\end{keyword}

\end{frontmatter}


\section{Introduction}
\label{Sec:Intro}

Mercury is a small planet (0.055 terrestrial mass) closest to
the Sun, which possesses the largest eccentricity (0.206) and
orbital inclination to ecliptic (over 7~degrees) among
the 8~planets of the Solar system \citep{all76}. Mercury has
also a global (but very weak) magnetic field and has almost
no atmosphere. All this properties as well as its proximity
to the Sun attract a special interest to the interaction of
Mercury with the solar wind, the model of the respective
ambient flow, and the possibility of survival of
the hermean trace (or ``shadow'') in the interplanetary medium in
the course of propagation of the solar wind (SW). The weak hermean
magnetosphere as well as the absence of an appreciable atmosphere
and ionosphere enable the enhanced solar wind to actually approach
the planetary surface, as it takes place in the interaction of
SW with Moon \citep{kab00,fuj07}. It is the aim of the present
paper to study the interaction of SW with Mercury by using
the variation in the SW velocity in the Earth's neighborhood
during inferior conjunctions, i.e., at the distance of 1~AU
from the Sun.

The solar wind is a quasi-steady radial supersonic outflow of
the solar coronal plasma through the entire Solar system;
its velocity being a few hundreds km/s, and the density,
about a few particles in a cubic cm. In this process, SW carries
the ``frozen'' magnetic fields, which interact with the planetary
magnetospheres.

SW velocity is one of the most difficult parameters for a reliable
forecast. Even in the clearest (most evident) cases, the amplitude
and duration of the increase after a flare will depend on many
parameters, such as a magnitude of the flare, its position on
the solar disk, magnetic field structure in the active region
and interplanetary space, direction of the outburst, its solid
angle, \textit{etc.} Therefore, the introduction of one more
parameter affecting SW velocity should enable us to get
an improved forecast.

\section{Material and Methods}
\label{Sec:MatMet}

To reveal the effect mentioned above, the mean daily velocities
were analyzed by the superimposed epoch method for the periods of
inferior conjunctions in 1995 to 2012.
The superimposed epoch method (or the ``synchronous detection'')
is based on the separation of the time interval under consideration
into the equal subintervals of the specified duration~$T$; and then
the observational data in the same temporal points from
the beginning of the above-mentioned subintervals are summed up or
averaged. As a result, if the data vary with~$T$ or a comparable
period, the corresponding effect is accumulated; while the data
varying in time randomly or with the periods incomparable with~$T$
are smoothed out.

The mean daily velocities were derived from the mean hourly ones,
which were measured by \textit{ACE} satellite in L1 Lagrangian point
in the vicinity of the Earth \citep{ace13}.
(Such a location of \textit{ACE} in the L1 point between the Earth
and Sun enables one to perform a continuous measurement of
the SW parameters and gives approximately an one-hour-advance
warning of the impending geomagnetic activity.)
At last, let us mention that differences in the space weather and
in the relative positions of the planets and Sun above ecliptic
and with respect to the perihelion of hermean orbit were not
taken into account in our analysis.

\section{Results}
\label{Sec:Res}

\begin{figure}
\includegraphics[width=13cm]{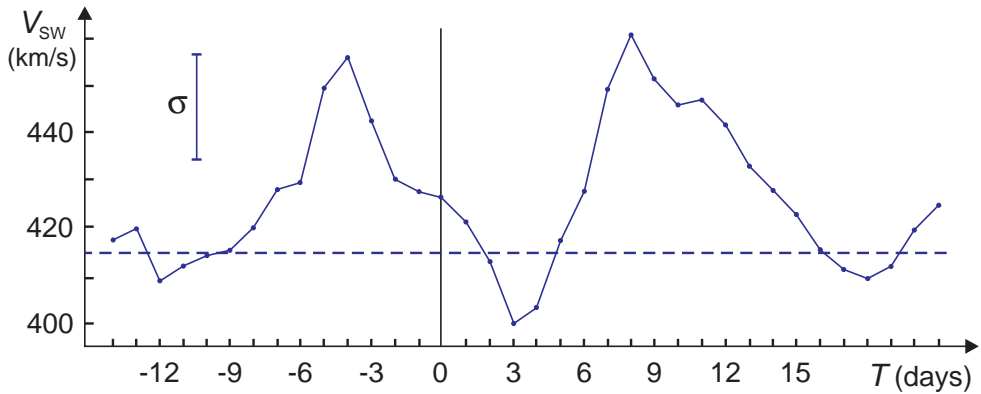}
\caption{\label{fig:1}
Mean variation of velocity of the solar wind during the inferior
conjunction with Mercury. Vertical line is the instant of
conjunction, horizontal dashed line is the mean velocity of
the solar wind in the interval under consideration (1995--2012).}
\end{figure}

Despite of the restrictions outlined in the previous section,
analysis of 54~oppositions (inferior conjunctions), presented
in Fig.~\ref{fig:1}, shows a noticeable increase in the SW velocity
within 4~days before the oppositions ($\sim$10\%) and 8~days
after them ($\sim$12\%) along with a decrease in the velocity
($\sim$4\%) within~3--4 days after the oppositions (which is
the Mercury's shadow). Such a delay corresponds to the average
SW velocity about 300--400~km/s, at which the radial SW passes
the distance from Mercury to the Earth.

\begin{figure}
\includegraphics[width=13cm]{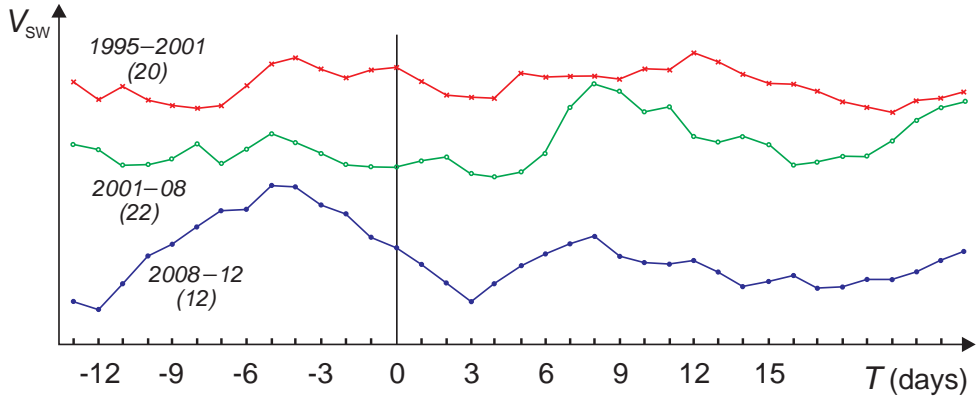}
\caption{\label{fig:2}
Characteristic features in variations of SW velocity in different
phases of the solar cycle. Upper curve refers to 1995--2001;
middle curve, 2001--2008; lower curve, 2008--2012. The values in
parentheses represent the numbers of inferior conjunctions in
each data set.}
\end{figure}

To verify this effect and to reveal its probable dependence on
the solar cycle phase, we have performed the same processing
separately for 3~data sets: from the minimum to maximum of
the 23rd cycle, for the period of its decay, and for the minimum
and the onset of the new 24th cycle (Fig.~\ref{fig:2}).
The above-mentioned characteristic features of the velocity
variations were found to survive in all 3~periods, although
the ratio of velocities before and after the oppositions
as well as their temporal localization with respect to
the conjunctions slightly varied.

These general tendencies are less expressed at the stage of
the cycle increase (1995--2001), most probably, because of
the weakness of the solar wind in the respective period.
Nevertheless, a position of the velocity minimum remains
the same, \textit{i.e.}, it takes place in the third or fourth
day after the opposition; the average SW velocity being 415~km/s
for the entire sample of 54~intervals of oppositions.

\section{Discussion and Conclusions}
\label{Sec:DisCon}

The results presented demonstrate that the trace of a planet in
the form of SW variation survives in the course of its propagation
in the interplanetary space up to the distances of, at least,
0.6~AU, \textit{i.e.}, about 90\,000\,000~km, which can hardly
remind a hydrodynamic flow around the ball.
The presence of even a relatively weak hermean magnetic field
evidently makes the flow of the solar wind around the planet more
regular and results in its laminar behavior.

\begin{figure}
\includegraphics[width=11cm]{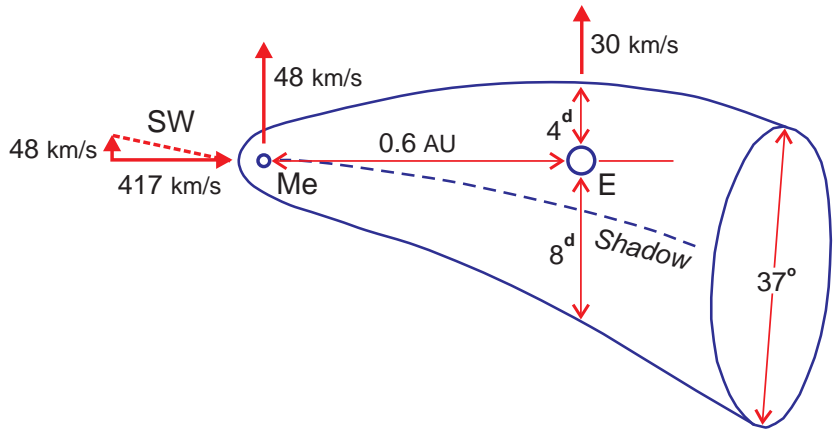}
\caption{\label{fig:3}
The model of the solar wind flow around Mercury.
SW is the solar wind, Me is Mercury, E is the Earth,
dashed curve is the Mercury's shadow.}
\end{figure}

The model of the flow of the solar wind around Mercury,
which takes into account the approximate axial symmetry of the
process and the excess of SW velocity by an order of magnitude
over the hermean orbital velocity, is shown in Fig.~\ref{fig:3}.
The perturbation of SW velocity in space looks like a conical
surface, similar to the paraboloid of revolution \citep{erk06};
the Mercury being located, roughly speaking, in its focus,
and the solid angle equaling about~37${}^{\circ}$
(12~days $\times$ 360${}^{\circ}$/116~days).
The Mercury's ``shadow'' goes approximately along the axis of
the cone of the SW velocity fluctuation.

It seems that the revealed nonuniformity of the solar wind
can be treated as a tail of the hermean magnetosphere, extending,
at least, up to the Earth. Such size of the tail substantially
exceeds the value 2\,500\,000~km, which was derived from
the sodium emission in the work by \citet{bau08}.

It would be very interesting to compare the features established
for Mercury with the flow of the solar wind around Venus,
which does not have a global magnetic field (as distinct from
Mercury) but possesses an extremely dense atmosphere.

It is reasonable to assume also that the same phenomena may
take place in the solar wind flow around the Earth, and
their consequences can be observed on the outer planets or
by the remote spacecraft.

\section*{Acknowledgements}

I am grateful to Yu.V.~Dumin for help in the translation and
preparation of the present paper.

This work was partially supported by the Russian Foundation
for Basic Research, grant no.~11-02-00843a.




\end{document}